\documentstyle[12pt,aaspp4]{article}
\begin{document}

\title{Galactocentric Distance With the OGLE and Hipparcos Red Clump Stars}

\author{Bohdan Paczy\'nski}
\affil{Princeton University Observatory, Princeton, NJ 08544--1001}
\affil{e-mail: bp@astro.princeton.edu}
\author{Krzysztof Z. Stanek}
\affil{Harvard-Smithsonian Center for Astrophysics, 60 Garden St., MS20,
Cambridge, MA 02174}
\affil{e-mail: kstanek@cfa.harvard.edu}

\begin{abstract}

We compare red clump stars with parallaxes known to better than 10\%
in the Hipparcos catalog and corrected for the interstellar
extinction, with the OGLE red clump stars in Baade's Window also
corrected for the interstellar extinction.  There are $\sim 600$ and
$\sim 10,000$ such stars in the two data sets, respectively.  We find
empirically that the average I-band magnitude of red clump stars does
not depend on their intrinsic color in the range $ 0.8 < (V-I)_0 < 1.4
$.  The red clump luminosity function is well represented by a
gaussian with the peak at $ M_{I_0,m} = -0.26 $, and the dispersion $
\sigma _{RC} \approx 0.2 $ mag.  This allows a single step
determination of the distance to the galactic center and gives $ R_0 =
8.4 \pm 0.4 $ kpc.  The number of red clump stars is so large that a
formal statistical error is only $\sim 1\% $.

The local stars are relatively blue and have a small color dispersion:
$\langle (V-I) \rangle = 1.01 $, $ \sigma _{(V-I)} = 0.08 $, while for
the bulge stars $\langle (V-I)_0 \rangle = 1.22 $, $ \sigma _{(V-I)_0}
= 0.14$.  Presumably, the bulge population has a broader range and a
higher average metallicity than the local disk population.

\end{abstract}

\keywords{
galaxy: center --
galaxy: fundamental parameters --
galaxy: solar neighborhood --
galaxy: stellar content --
stars: horizontal-branch
}

%\section 1
\section{Introduction}

The center of our Galaxy is at $R_0 \approx 8.0 \pm 0.5$ kpc (Reid
1993).  The Hipparcos catalogue of parallaxes (Perryman et al.~1997)
will improve this distance measurement.  The purpose of this paper is
to present an estimate based on the comparison between the red clump
giants measured by Hipparcos and by OGLE (Optical Gravitational
Lensing Experiment, cf. Udalski et al. 1992).  These stars are the
metal rich equivalent of the better known horizontal branch stars, and
theoretical models predict that their absolute luminosity only weakly
depends on their age and chemical composition (Seidel, Demarque, \&
Weinberg 1989; Castellani, Chieffi, \& Straniero 1992; Jimenez, Flynn,
\& Kotoneva 1997).  Indeed the OGLE color-magnitude diagram of stars
in Baade's Window (Udalski et al.~1993; Paczy\'nski et al.~1994, their
Fig.1; Kiraga, Paczy\'nski \& Stanek 1997, their Fig.6), and the
Hipparcos' absolute magnitude-color diagram (Perryman et al.~1997,
their Fig.3) clearly show how compact the red clump is.  In this paper
we determine the variance in the absolute I band magnitude to be only
$ \sim 0.2 $ mag.

Any method of the galactocentric distance determination which is based
on stars suffers from at least four problems:

\begin{enumerate}

\item The accuracy depends on the absolute magnitude determination for
the nearby stars.

\item Interstellar extinction has to be determined for the stars near
the galactic center as well as for those near the sun.

\item The masses, ages, and chemical composition may be different for
the stars near the galactic center and for their counterparts near the
sun.

\item The statistical error is large if the number of objects is small.

\end{enumerate}

The red clump giants are the only type of stars which do not suffer
from the fourth problem.  In spite of their large number and sound
theoretical understanding these stars have seldom been used for the
galactic structure studies.  However, recently Stanek (1995) and
Stanek et al.~(1994, 1997) used these stars to map the Galactic bar,
and also to map the interstellar extinction in Baade's Window
(Wo\'zniak \& Stanek 1996; Stanek 1996).  In this paper we compare the
absolute magnitudes of $\sim 600$ nearby red clump stars with accurate
(better than 10\%) trigonometric parallaxes measured by Hipparcos with
the apparent magnitudes of the red clump stars measured by the OGLE in
Baade's Window.  This comparison gives the galactocentric distance in
a single step.  In the last section we discuss various effects which
have to be evaluated in order to reduce the systematic error.

%\section 2
\section{The method and preliminary result}

Inspection of the color-magnitude diagrams based on Hipparcos and OGLE
data reveal a strong dependence of the visual magnitude of red clump
giants on their color, while their $I$-band magnitudes reveal no color
dependence.  Thus, on purely observational grounds, the $I$-band turns
out to be the best in the applications in which the red clump stars
are used as standard candles.  It is possible that bolometric
corrections to the $I$-band are very small for these moderately cool
stars, and theoretical models show only weak dependence of $M_{bol}$
on either age or chemical composition (Seidel, Demarque, \& Weinberg
1989; Castellani, Chieffi, \& Straniero 1992; Jimenez, Flynn, \&
Kotoneva 1997).

Using the extinction map of Stanek (1996) with the zero point
determined by Gould, Popowski \& Terndrup (1997) and by Alcock et
al.~(1997) we selected the region of Baade's Window with $A_V<1.5$, to
minimize the effects of a possible error in the ratio $A_V/E_{V-I}$.
The $I_0-(V-I)_0$ color-magnitude diagram for this part of Baade's
Window is shown in Fig.1.  We divided the red clump region into five
color bins and we fitted the observed distribution of the stars in
each bin with a function
\begin{equation}
n(I_0) = a + b (I_0-I_{0,m}) + 
\frac{N_{RC}}{\sigma_{RC}\sqrt{2\pi}}
 \exp\left[-\left(\frac{I_0-I_{0,m}}{2\sigma_{RC}}\right)^2 \;\right],
\end{equation}
where the first two terms describe a fit to the ``background''
distribution of the red giant stars, and the gaussian term represents
a fit to the red clump itself.  The locations of five points with the
vertical error bars in Fig.1 correspond to the values of $I_{0,m}$ and
their formal $3\sigma$ errors for the five color bins.  No
statistically significant color trend is apparent in $I_{0,m}$.

\begin{figure}[t]
\plotfiddle{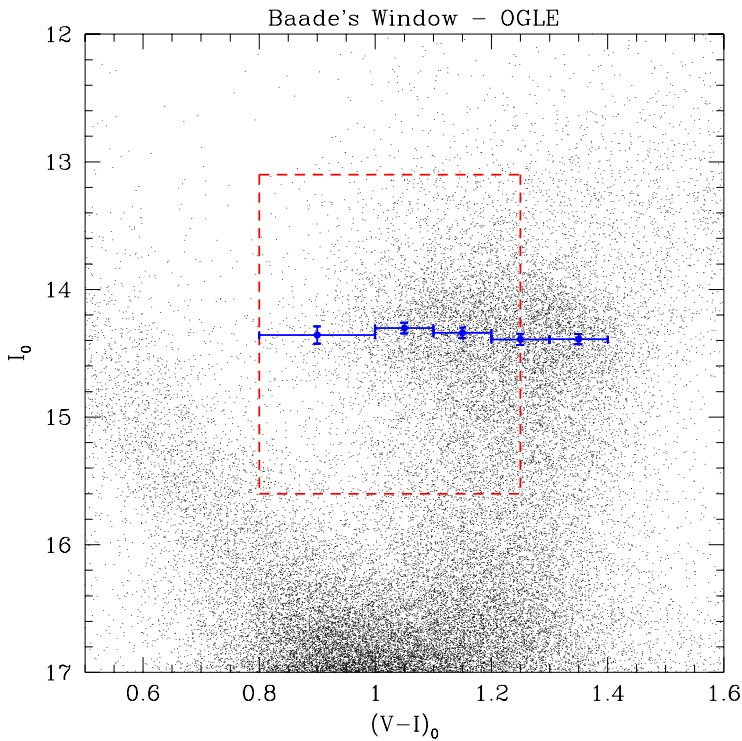}{8cm}{0}{50}{50}{-160}{-90}
\caption{The color-magnitude diagram corrected for interstellar
extinction is shown for the stars in the part of Baade's Window with a
small extinction, $ A_V < 1.5$ mag.  The dashed rectangle surrounds
the red clump region used for the comparison with the local stars.
The ridge of the highest density of the red clump stars, $ I_{0,m} $,
is shown for five color bins in the range $ 0.8 < (V-I)_0 < 1.4$, as
calculated with the Eq.1.  It is apparent that the value of $I_{0,m}$
does not depend on the $(V-I)$ color.}
\end{figure}

\begin{figure}[t]
\plotfiddle{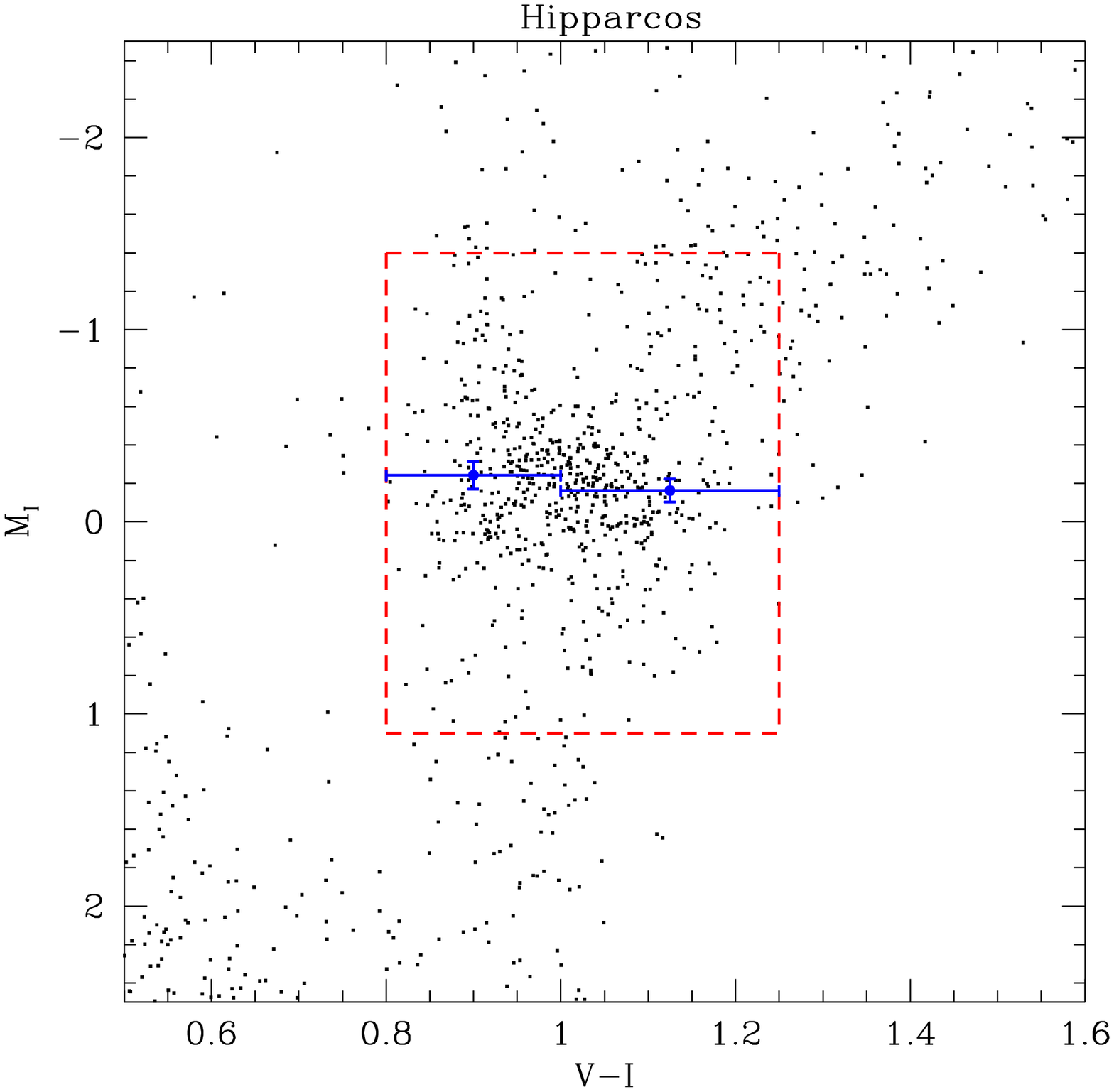}{8cm}{0}{50}{50}{-160}{-90}
\caption{The color-magnitude diagram is shown for the Hipparcos stars
with trigonometric parallax errors smaller than 10\%.  The dashed
rectangle surrounds the red clump region used for the comparison with
Baade's Window stars.  The ridge of the highest density of the red
clump stars, $M_{I,m}$, is shown for two color bins in the range $0.8
< (V-I) < 1.2$, as calculated with the Eq.1.  Note that the average
red clump color is $ \langle (V-I) \rangle \approx1.0$, compared to
the average value in Baade's Window $ \langle (V-I)_0 \rangle
\approx1.2$, indicating that the galactic bulge stars are
substantially more metal rich than the solar neighborhood.  Hipparcos
$ (V-I) $ colors are given to two decimal places only.  This figure
was made with a random third decimal digit added to the data to avoid
the appearance of artificial bands.}
\end{figure}

The Hipparcos based absolute magnitude-color diagram is shown in Fig.2
for the stars with parallaxes measured to better than 10\%.  The
Cousins' $ (V-I) $ colors in the catalog come from various sources.
For this study we used only those stars for which the source of
$(V-I)$ was labeled A, C, E, F, or G, i.e.  those for which I-band
photometry was available.  There are 664 such stars within the dashed
rectangle in Fig.2.  We divided them into two color bins and fitted
their distribution with the formula analogous to that given in Eq.1,
with the $I_0$ replaced with the $M_I$. There is no significant color
trend in $M_I$.

Given the fact that the average luminosity of the red clump stars
appears to be independent of their color in Baade's Window (Fig.1) and
in the solar neighborhood (Fig.2), it seems safe to determine the
distance to the galactic center by comparing these two populations.
It is also very fortunate that even though the bulk of the bulge
population is redder, and hence it seems to be more metal rich than
the solar neighborhood, there is a full overlap in the colors.  In
this first approximation we neglect the extinction of the Hipparcos
red clump giants as their average distance is only $87\;pc$.  We
selected the red clump stars in the color range $0.8<(V-I)_0<1.25$ in
Baade's Window (there were 7174 of them) and $0.8 < (V-I)< 1.25$ in
Hipparcos catalog, and we fitted both distributions with Eq.1.  We
obtained $I_{0,m}=14.323\pm0.009$ and $M_{I,m} = -0.185\pm0.016$.
Therefore the distance modulus is $(m-M)_{BW} = ( I_{0,m} - M_{I,m} )
= 14.508\pm0.019$, and $d_{BW}=7.97\;kpc$, with the formal statistical
error of only $\pm80\;pc$.

\begin{figure}[t]
\plotfiddle{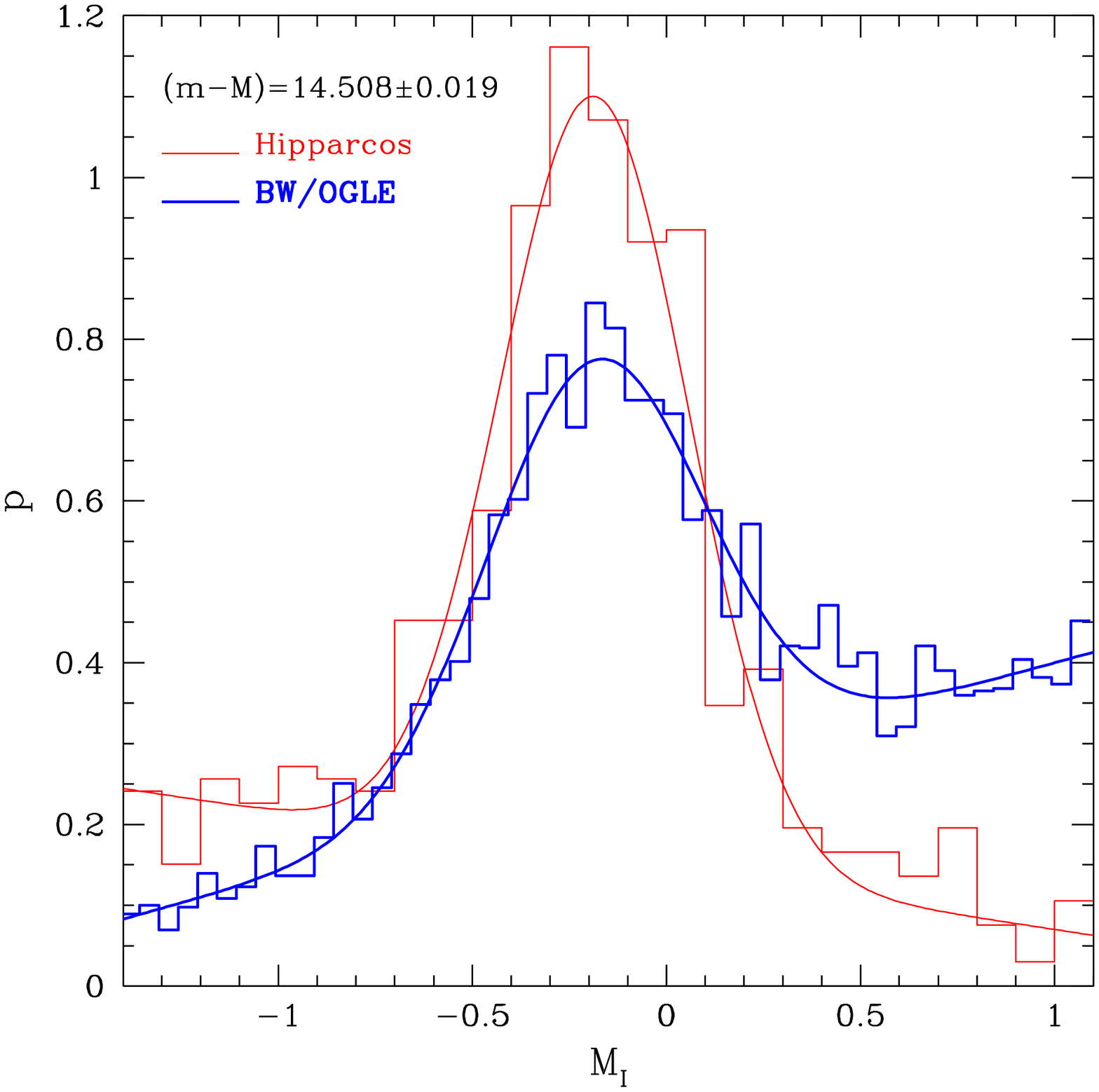}{8cm}{0}{50}{50}{-160}{-90}
\caption{The number of red clump stars in the solar neighborhood,
based on Hipparcos data, is shown as a function of absolute magnitude
$M_I$ with the thin solid line (fit and histogram). The number of red
clump stars in Baade's Window, based on the OGLE data, is shown as a
function of absolute magnitude $M_I$ with thick solid line, adopting
the preliminary distance modulus: $I_0 - M_I = 14.508$.  All
distributions are normalized.}
\end{figure}

The distribution of the two groups of the red clump stars as a
function of their absolute I-band magnitude is shown in Fig.3 with two
histograms as well as with two analytical fits of the type described
by the Eq.1.  All these distributions are normalized.  The gaussian
fitted to the OGLE red clump distribution has $\sigma_{RC}=0.28\;mag$
and the gaussian fitted to the Hipparcos distribution has
$\sigma_{RC}=0.24\;mag$.

The distribution of the red clump stars as a function of their $(V-I)$
color is shown in Fig.4.  It is clear that the bulge population is
redder, and presumably more metal rich than the nearby disk
population.

\begin{figure}[t]
\plotfiddle{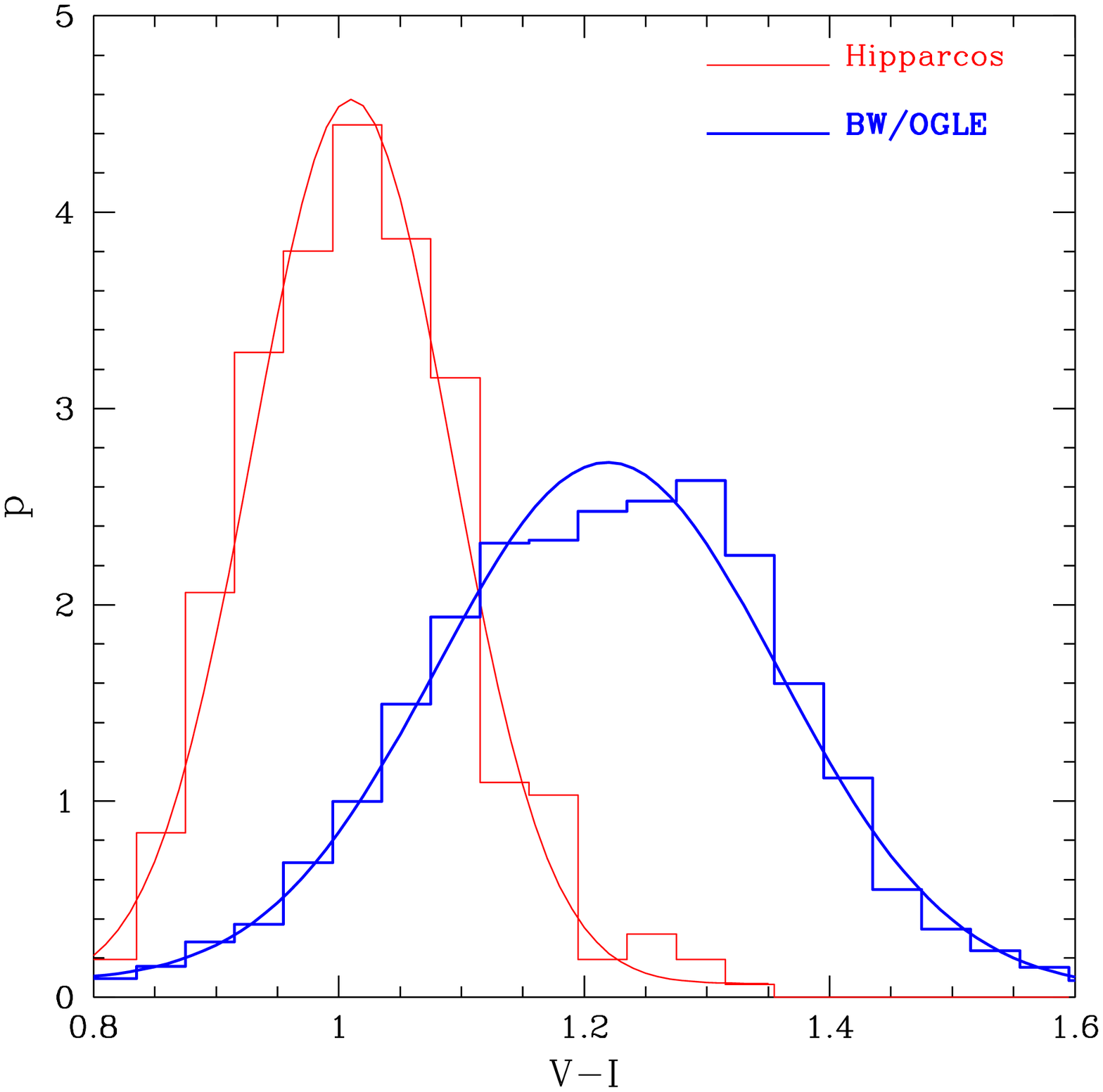}{8cm}{0}{50}{50}{-160}{-90}
\caption{The number of red clump stars in the solar neighborhood,
based on Hipparcos data, is shown as a function of their $(V-I)$ color
with the thin solid line. The number of red clump stars in Baade's
Window, based on the OGLE data, is shown as a function of their
$(V-I)_0$ color with the thick solid line.  All distributions are
normalized, and include all stars with their absolute magnitude in the
range: $ -0.485 < M_I < 0.115 $, i.e.  within $0.3;mag$ of $M_{I,m}$.
The bulge stars are much redder and presumably more metal rich than
nearby disk stars.}
\end{figure}

%\section 3
\section{Correcting some errors}

The very small formal statistical error for our galactocentric
distance follows from the very large number of red clump stars
measured by Hipparcos and OGLE.  The number of these stars is several
orders of magnitude larger than the number of either RR Lyrae or Mira
variables, and the observed dispersion in their magnitudes is only $
\sigma_{RC}\approx0.24\;mag$.  However, in addition to very small
statistical errors there are much larger systematic errors, some of
which we try to estimate and correct for.

We checked if the 10\% error limit adopted by us was not too generous,
and we repeated the analysis with a more stringent 5\% upper limit to
the parallax errors.  This reduced the number of stars within the
rectangle shown in Fig.2 from 664 to 240, and gave $M_{I,m} = -0.173
\pm 0.027$, well within small statistical errors of $M_{I,m} = -0.185
\pm 0.016$ obtained for the 10\% sample.  The dispersions were also
similar: $\sigma_{RC}=0.243$ mag for the 10\% sample and
$\sigma_{RC}=0.225$ mag for the 5\% sample.

By selecting stars with accurate parallaxes from the Hipparcos catalog
we introduced a distance bias which depends on the absolute magnitude:
the stars which are intrinsically brighter can be accurately measured
out to a larger distance.  Therefore, there are relatively more bright
stars in our Hipparcos sample than in our OGLE sample, as it is
apparent in Fig.3.  The difference in the slope of the red giant
``background'', i.e. the different value of the `b' coefficient in
Eq.1, introduces a shift in the location of the gaussian peak
describing the red clump.  In order to correct for the differences in
the two samples we multiplied the number of Hipparcos stars in each
magnitude bin by the factor
\begin{equation}
f \sim \frac{a_{BW} + b_{BW} M_I}{a_H + b_H M_I}.
\end{equation}
Following this procedure the new normalized fit to the modified
Hipparcos histogram gave new values of the $(a_H,b_H)$ coefficients
consistent with their Baade's Window values, and also a new value of
the gaussian peak: $M_{I,m} = -0.125\pm0.019$, which differs by
$0.06\;{\rm mag}$ from our original estimate.

In order to estimate the effect of interstellar extinction we selected
a sub-sample of 228 Hipparcos stars with the distance $d<70$ pc, and
we determined the parameters of the best fit to the luminosity
function described with Eq.1, correcting for the (now reduced)
distance bias in the way described in the previous paragraph.  We
found that $M_{I,m} = -0.192 \pm 0.023$, and $\sigma_{RC}=0.208$ mag
for these nearby stars, with the average distance $\langle d_{<70}
\rangle = 50\;pc$.  For the remaining 437 Hipparcos red clump giants
with the distance $d>70\;pc$ we obtained $M_{I,m} = -0.094 \pm 0.027$,
$\sigma_{RC}=0.234$ mag and the average distance $\langle d_{>70}
\rangle = 106\;pc$. The more distant stars are fainter by 0.098 mag,
with the difference in average distance being $56\;pc$.  The most
obvious explanation for this difference is the interstellar
extinction.  Extrapolating these results to the distance $d=0$, and
therefore to zero extinction, we obtain $M_{I_0,m} = -0.279$.  Our
final estimate of the correct average magnitude of the nearby red
clump stars is 0.088 mag brighter than our original crude estimate.
We adopt this difference as our crude estimate of the possible
systematic error in the absolute magnitude determination of the nearby
red clump giants.

The centroid of the gaussian distribution of red clump giants in
Baade's Window corresponds to the distance which differs from the
galactocentric distance for two reasons.  First, because of bar
asymmetry the point with the highest space density of the red clump
stars seen in the direction $ (l,b) = (1,-4)\;\deg$ is somewhat closer
than the galactic center.  Second, the volume within the observed
solid angle increases with the distance, which implies that we see
more stars on the far side than on the near side of the bar, and the
observed centroid is shifted to a somewhat larger distance than the
location of the highest space density of the bar stars.  We used the
E2 bar model of Stanek et al.~(1997) to calculate the two opposing
effects, and we found that the galactocentric distance modulus is
larger by $0.02\;{\rm mag}$ than the distance modulus obtained by us
for the apparent centroid of the star distribution in Baade's Window.
Therefore, the apparent magnitude of red clump giants at the galactic
center distance, corrected for reddening, is $I_{0,m} = 14.343$ mag,
with the error due to our model estimated at 0.02 mag.  A larger
contribution to the error comes from zero point in the interstellar
extinction, estimated to be $ \pm 0.05 $ mag (Gould, Popowski \&
Terndrup 1997; Alcock et al. 1997).

Combining the corrected values of the absolute magnitudes of the
nearby red clump stars, $M_{I_0,m} = -0.279 \pm 0.088$, with the
apparent magnitude of the red clump stars at the galactocentric
distance, $I_{0,m} = 14.343 \pm 0.05$, we obtain the corrected
distance modulus to the galactic center: $(M-m)_{GC} = 14.62 \pm
0.10$, and the corresponding distance $R_0 = 8.4 \pm 0.4\;kpc$.

There is yet another type of a systematic error possible: the age, the
chemical composition, and the masses of red clump giants may be
systematically different in the bulge/bar and near the sun.  The
apparent systematic difference in the colors implies that the two
populations are in fact different.  Recent stellar evolutionary models
(Jimenez et al. 1997) indicate that in the age range $ 2 - 12 $
billion years the effective temperature is dominated by the
metallicity.  This suggests that the bulge population is on average
more metal rich than the solar region.  This seems to be in conflict
with McWilliam \& Rich (1994), who found that giants in the bulge and
locally have similar metallicity.  It also seems that among the bulge
and local giants the most extreme stars are about equally super
metal-rich (Castro et al. 1996).

It is not clear at this time what is the reason for the major
difference in colors between the bulge and the local clump giants, but
at least two possibilities may be considered.  The spectroscopic
samples are relatively small, and the selection criteria are not
clear.  On the other hand, the photometric samples are more complete,
or at least they should have no color bias.  It is possible, that
while the range of metallicity in the bulge and near the sun is
similar, the distributions are different, with the super metal-rich
stars more common in the bulge.  Another possibility is that
photometry and spectroscopy refer to different metallicities.  The $
(V-I) $ color is determined mostly by the effective temperature, which
in turn depends on the photospheric continuum opacity in stars with
deep convective envelopes, such as the red clump giants.  It is not
obvious that the spectroscopic metallicity based on line opacities is
weighted in the same way as the photometric metallicity based on
continuum opacity.  In any case, the issue of color difference has to
be resolved for the distance estimate based on the clump giants to be
trusted.

%\section 4
\section{Discussion}

We found empirically that the average absolute magnitude of the red
clump stars as measured in the Cousin's I-band is independent on their
$(V-I)$ color.  It is important to reproduce this property with
theoretical models, to have a better assessment of the robustness of
our assertion, i.e. to have theoretical values of $M_I$ calculated for
a broad range of stellar ages and chemical compositions.  Note that
stellar populations are different in the galactic bulge and in the
solar neighborhood, and a theoretical understanding of the
$M_{I,m}-(V-I)$ relation is highly desirable.

Among the various stellar distance indicators the red clump giants are
likely to be the best for determining the galactocentric distance
because there are so many of them.  In particular Hipparcos provided
accurate distance determinations for almost 2,000 such stars, but
unfortunately I-band photometry is available for only $\sim 30\%$ of
them.  It is important to obtain I-band photometry for all Hipparcos
red clump giants in order to make a better correction for at least two
effects: the distance bias in the catalog and the interstellar
extinction.  It would be useful to make accurate I-band photometry for
all $\sim 10^6$ stars in the Tycho catalogue (H${\rm \o}$g et
al.~1997).  Another important work which will be done soon with the
forthcoming OGLE-2 $BVI$ photometry of $\sim 3 \times 10^7$ stars in
the bulge (Udalski, Kubiak \& Szyma\'nski 1997), is a better bulge/bar
model and a better extinction map of that region.  This will reduce
the uncertainty in the OGLE/Hipparcos comparison.

\acknowledgments{It is a great pleasure to acknowledge helpful
suggestions by Andy Gould, Michael Richmond, Michael Strauss and
Przemek Wozniak, and the anonymous referee.  This work was supported
by the NSF grants AST--9313620 and AST--9530478.  KZS was supported by
the Harvard-Smithsonian Center for Astrophysics Fellowship.}

%\newpage

%REFERENCES

\end{document}